\documentclass{aastex61}
\shorttitle{SU Lyn: diagnosing the boundary layer with UV and hard X-ray data}
\shortauthors{R. Lopes de Oliveira et al.}

\begin{document}

\title{SU Lyn: diagnosing the boundary layer with UV and hard X-ray data}

\correspondingauthor{R. Lopes de Oliveira}
\email{raimundo.lopesdeoliveirafilho@nasa.gov, rlopes@ufs.br}

\author{R. Lopes de Oliveira}
\affiliation{X-ray Astrophysics Laboratory, NASA Goddard Space Flight Center, Greenbelt, MD 20771, USA
}
\affiliation{Departamento de F\'isica, Universidade Federal de Sergipe, Av. Marechal Rondon, S/N, 49000-000, S\~ao Crist\'ov\~ao, SE, Brazil
}
\affiliation{Department of Physics, University of Maryland, Baltimore County, 1000 Hilltop Circle, Baltimore, MD 21250, USA}
\author{J. L. Sokoloski}
\affiliation{Columbia Astrophysics Lab 550 W120th St., 1027 Pupin Hall, MC 5247 Columbia University, New York, New York 10027, USA}
\author{G. J. M. Luna}
\affiliation{CONICET-Universidad de Buenos Aires, Instituto de Astronom\'ia y F\'isica del Espacio, (IAFE), Av. Inte. G\"uiraldes 2620, C1428ZAA, Buenos Aires, Argentina}
\affiliation{Universidad de Buenos Aires, Facultad de Ciencias Exactas y Naturales, Buenos Aires, Argentina}
\affiliation{Universidad Nacional Arturo Jauretche, Av. Calchaqu\'i 6200, F. Varela, Buenos Aires, Argentina}
\author{K. Mukai}
\affiliation{CRESST and X-ray Astrophysics Laboratory, NASA Goddard Space Flight Center, Greenbelt, MD 20771, USA
}
\affiliation{Department of Physics, University of Maryland, Baltimore County, 1000 Hilltop Circle, Baltimore, MD 21250, USA}
\author{T. Nelson}
\affiliation{Department of Physics and Astronomy, University of Pittsburgh, Pittsburgh, PA 15260, USA}

\begin{abstract}

Symbiotic stars in which the symbiotic phenomenon is powered solely
by accretion, often at an average rate that is higher than in
cataclysmic variable stars, provide an important opportunity to
diagnose boundary layers around disk-accreting white dwarfs.
Here we investigate SU\,Lyncis, a recently discovered example of
a purely accretion-powered symbiotic star, using the first reliable
X-ray spectroscopy, obtained with NuSTAR, and UV photometry obtained
with Swift. SU\,Lyn has hard, thermal, X-ray emission that is
strongly affected by a variable local absorber -- that has little 
impact on the UV emission. Its X-ray spectrum
is described well using a plasma cooling from $k$T\,$\approx$\,21\,keV, with
a 3 to 30 keV luminosity of approximately 4.9$\times$10$^{32}$\,ergs\ s$^{-1}$.
The spectrum is also consistent with the presence of reflection with an
amplitude of 1.0, although in that case, the best-fit plasma temperature is 20-25\% lower.
The UV to X-ray luminosity ratio of SU Lyn changed significantly
between 2015 and 2016. We interpret this as a consequence of a 
drop by almost 90\% in the accretion rate. Whereas the UV luminosity of the
disk responded linearly, the luminosity of the optically thin (hard X-ray)
emission from the boundary layer remained roughly constant because the
boundary layer changed from partially optically thick to almost completely
optically thin.  Under this interpretation, we place a lower limit on the
white dwarf mass of 0.7\,M$_{\odot}$ (0.8\,M$_{\odot}$ if we neglect reflection).

\end{abstract}


\section{Introduction}

Phenomenologically, symbiotic stellar binaries were initially defined by the presence of strong high-excitation emission lines in the optical on top of a red giant continuum \citep{1986syst.book.....K}. They are associated with binary systems involving a red giant star and an accretor that might be a white dwarf or a neutron star \citep[see][for a review]{2017PASP..129f2001M}.  
Most members of the class have been discovered in the optical using the strong high-excitation
lines as the defining characteristic.

What may be just the tip of an iceberg came from the discovery by \citet{2016MNRAS.461L...1M} that the red giant SU Lyncis (SU Lyn) is the optical counterpart of a hard, thermal X-ray source. Its properties, including excess in UV when compared to non-interacting red giants and variability in optical lines of Hydrogen Balmer series, [NeIII], and Ca II, are consistent with accretion onto a white dwarf without shell burning.
The hard X-ray nature of the system was firstly identified by the authors from the Swift/BAT hard X-ray all-sky survey, then followed by a coordinated follow up in X-rays and UV conducted with the Swift satelite (XRT and UVOT cameras) and in the optical through medium and high-resolution spectroscopy from two telescopes at Asiago. Altogether, the observations supported that the X-ray emission is dominated by an optically thin plasma that can be as hot as 2$\times$10$^{8}$ K ($k$T of about 17 keV; from the \textsc{apec} model) -- or peaking at 3$\times$10$^{8}$ K when assuming a cooling flow model (\textsc{mkcflow}).  These properties are reminiscent of the $\delta$-type symbiotic stars, which is currently composed of about a dozen members and as defined by \citet{2013A&A...559A...6L}: ``{\it highly absorbed, hard X-ray sources [...]. The likely origin is the boundary layer between an accretion disk and the white dwarf}''. SU Lyn is a long-term variable X-ray emitter in both soft and hard X-rays, strongly affected by local absorbers that change with time. Besides revealing the presence of weak high-ionization lines, the optical observations led \citet{2016MNRAS.461L...1M} to conclude that SU Lyn is an M5.8III cool giant star located at $d$\,=\,640$\pm$100 pc, and finally pointing out that it is a member of a symbiotic system.

The potential significance of this discovery lies in the fact that there may be a large population
of symbiotic stars with weak emission lines. Because of this, this population has remained
hidden, only to be revealed by their high energy emission. The number of symbiotic stars
in the Galaxy and their contribution to the integrated X-ray emission have likely been
underestimated. Moreover, this new subclass of symbiotic stars opens up a new avenue to
investigate accretion in, and evolution of, symbiotic stars. The prototype, SU Lyn, is likely
to be among the brightest member of the subclass and therefore deserves further attention.
In this paper, we present a more in depth investigation of SU Lyn in the X-ray and UV domains
using coordinated NuSTAR and Swift observations.

\section{Observations}

While \citet{2016MNRAS.461L...1M} investigated the hard X-ray properties of SU\,Lyn
using Swift BAT data, the sensitivity of this instrument is such that
it takes months of integration to securely detect this source (Mukai
et al. used 120 d bins in their Figure 1, showing the BAT light curve).
The best S/N, integrated over the high state in the 15-35 keV range,
is 16.3. Since we must divide these data into multiple energy bins
for spectral analysis, hard X-ray spectral parameters of SU Lyn cannot
be tightly constrained using Swift BAT. 

We therefore observed SU Lyn with the Nuclear Spectroscopic Telescope
Array \citep[NuSTAR;][]{2013ApJ...770..103H} mission. The total exposure time of
about 40.9\,ks, for both Focal Plane Modules A and B (FPMA and FPMB),
was spread over approximately 80 ks on 2016 August 12-13 (ObsID
30201025002; Table \ref{tbl:obs}). During about 9.4 ks out of that time, we also observed
it using the Neil Gehrels Swift Observatory \citep{2004ApJ...611.1005G}
using both X-Ray Telescope (XRT) and UltraViolet and Optical Telescope
(UVOT) instruments (ObsID 00081892001). We added to our analysis
the Swift/UVOT (ObsID 00034150001, taken on 2015 November 20) and BAT data 
of SU Lyn investigated by \citet{2016MNRAS.461L...1M} to perform
a comparative study of the state of the source in UV and X-rays.

We used the XRT in photon counting (PC) mode because the expected
X-ray flux was modest. However, the CCD detector for the XRT is
also sensitive to optical and near-infrared photons, and lead to
a phenomenon called optical loading
\footnote{http://www.swift.ac.uk/analysis/xrt/optical\_loading.php}.
When numerous optical/near IR photons land on the same pixel during a single CCD
exposure, the total amount of charge
resulting
from this can resemble that from a single, soft X-ray photon. This
can result in spurious events at soft energies. Such a soft component
is seen in the 2016 August XRT data on SU Lyn.

In addition, the on-board software uses the distribution of charges
across neighboring pixels to distinguish between X-ray events and
particle hits. True X-ray photons lead to a tightly clustered set
of charges (single pixel to a 2x2 square), while particle hits tend
to leave charges in multiple adjacent pixels. In the presence of optical
loading, X-ray photons can be misinterpreted as having particle
hit-like distribution of charges, and therefore be rejected in
standard screening. During both 2015 and 2016 observations of SU Lyn,
while the events in the unscreened files have a point spread function
(PSF) like spatial distribution, the cleaned files have a more diffuse
appearance with a central hole. We conclude that optical loading
led to the loss of true X-ray event, and distortion of pulse height
distribution of X-ray events that survive the screening. We therefore
did not use the XRT data in this paper, and also conclude that the X-ray
spectral parameters of SU Lyn derived from Swift/XRT data by \citet{2016MNRAS.461L...1M} 
are unreliable, thereby making the NuSTAR data the first reliable X-ray spectra
of SU Lyn.

Data reduction and analysis were carried out with \textsc{heasoft} version 6.22 using the specific tools for each mission and calibration files available in August 2017 and December 2017 for the NuSTAR and Swift data, respectively. For the NuSTAR data we applied the \textsc{nupipeline} and then the \textsc{nuproducts} to obtain science products, while for the Swift/UVOT data we used the tasks \textsc{coordinator}, \textsc{uvotscreen}, \textsc{xselect}, and \textsc{uvotevtlc}. The typical absolute astrometric uncertainty of the NuSTAR is $\pm$8$\arcsec$, with a usual relative offset of 5-10\arcsec \,for one NuSTAR module with respect to the other -- drifting on time depending essentially on the thermal condition and therefore on the illumination of NuSTAR by the Sun (Brian Grefenstette; priv. commun.). Thus, we use different extraction regions for FPMA and FMPB based on their individual images.

The spectral analysis was conducted with the X-ray spectral fitting package \textsc{xspec v12.9.1m}.
It was based on simultaneous fits of NuSTAR FPMA and FPMB data from the minimum detector threshold of 3\,keV \citep[][and references therein]{2015ApJS..220....8M} to 30 keV, 
since background dominates at higher energies.
The energy binning was set to have a minimum of 25 counts in each bin.
A multiplicative constant was applied for each dataset in \textsc{xspec} to account for possible cross-calibration uncertainties but the difference is limited to 5\%.
The BAT dataset was important to characterize the photometric variability of the system. All errors and comparisons discussed in this work are at a 1\,$\sigma$ confidence level.

\begin{table}
\begin{center}
\caption{2016 August observations of SU\,Lyn.\label{tbl:obs}}
\begin{tabular}{lcccc}
\tableline\tableline
             &  NuSTAR & Swift \\
\tableline
ObsID        & 30201025002             & 00081892001            \\
Start Time   & 2016-08-12 20:16:08     & 2016-08-12 22:28:58    \\
Stop Time    & 2016-08-13 17:36:08     & 2016-08-13 10:41:52    \\
Exposure     & 40920\,s                & 9411\,s             \\
\hline
\tableline
\end{tabular}
\end{center}
\end{table}

\section{Results}

\subsection{X-ray spectral energy distribution}
\label{sct:spct}

\begin{figure*}[h!]
\includegraphics[angle=-90,scale=.36]{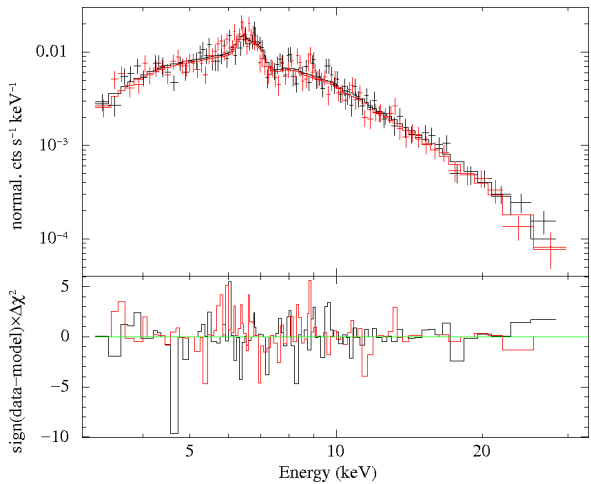}
\caption{NuSTAR X-ray spectra: FPMA in black and FPMB in red. Spectral fit corresponds to model 1 (M1; \textsc{tbabs*(apec+gauss))}.\label{fig:spct}}
\end{figure*}

We constructed two spectra from the whole NuSTAR observation, one from the FPMA and another from the FPMB data, then fitted simultaneously with \textsc{xspec} (Figure \ref{fig:spct}). 
Their shape and the clear presence of emission lines of the Fe\,K complex around 6.7\,keV reinforces three characteristics previously reported by \citet{2016MNRAS.461L...1M} that are investigated in detail in this work: hard, thermal, and locally absorbed X-ray emission. 

We adopt the \textsc{tbabs} model to account for the photoelectric absorption \citep{2000ApJ...542..914W}. Two models based on thermal plasma emission were individually applied to test two different hypotheses: the emission from a single-temperature plasma component (\textsc{apec}) 
and the emission from a cooling-flow plasma (\textsc{mkcflow}). The abundance table applied in the models is that of \citet{2000ApJ...542..914W}. For comparison, we find that the use of such a table results in mutually consistent temperatures ($k$Ts) but absorption columns (in equivalent Hydrogen column, N$_{H}$) and abundances that are systematically higher by 52\% and 75\%, respectively, with respect to the values obtained from Solar abundance vector set to \citet{1989GeCoA..53..197A}. In all cases, we added a Gaussian line (\textsc{gaussian}) to account for the excess due to the fluorescence Fe line at 6.4\,keV. As the line parameters are not well determined during the fit, the centroid was fixed at the rest energy value and the line width ($\sigma$) was fixed to 1\,eV, while the normalization was a free parameter during the fit.

Single-temperature plasma and cooling-flow spectral models both provided acceptable fits to the NuSTAR spectra.
We first attempted to fit the spectra with the simplest model, \textsc{tbabs*(apec+gauss)} (M1). We found that the spectrum of SU Lyn is well described by a plasma with temperature ($k$T) of 12.0$^{+0.8}_{-0.7}$\,keV and abundance ($Z$) of 0.90$^{+0.22}_{-0.19}$ relative to solar values ($Z_{\odot}$). The photoelectric absorption is equivalent to a Hydrogen column (N$_{H}$) of 19.8$^{+1.8}_{-1.8}$\,$\times\,$10$^{22}$\,cm$^{-2}$. 
The ionized lines of the Fe\,K complex are well described by M1. This fit results in $\chi^2_{\nu}$ equal to 1.09 for 182 degrees of freedom (d.o.f).

The other model, 
\textsc{tbabs*(mkcflow+gauss)} (M2), did not significantly improve the fit. 
M2 resulted in N$_{H}$\,=\,23.3$^{+2.1}_{-2.3}$\,$\times\,$10$^{22}$\,cm$^{-2}$ and $Z$\,=\,0.75$^{+0.20}_{-0.16}$\,$Z_{\odot}$. As the low energies are not covered by the NuSTAR spectra, we fixed the low temperature component of the \textsc{mkcflow} to its lower limit of $k$T\,=\,80.8\,eV. The maximum temperature parameter of this model has $k$T$_{H}$\,=\,21.1$^{+2.6}_{-1.9}$\,keV. The $\chi^2_{\nu}$ obtained from M2 is 1.07 for 182 degrees of freedom. 
For M2 we set the {\it switch} parameter to 2, such that the \textsc{mkcflow} model, originally based on the \textsc{mekal} code, was calculated by running the \textsc{apec} (AtomDB) table. We adopt a redshift of 1.49$\times$10$^{-7}$ for the \textsc{mkcflow} component of M2 and the default cosmology parameters in \textsc{xspec} (H$_{0}$ = 70 q$_{0}$ = 0, and $\Lambda_{0}$ = 0.73) -- to take into account the distance of the system, which \citet{2016MNRAS.461L...1M} estimated to be 640\,pc \citep[consistent with the Gaia DR2 parallax of $1.493 \pm 0.096$\,mas;][]{2016A&A...595A...1G,2018arXiv180409365G}. We tested the inclusion of a partial covering fraction absorption (\textsc{pcfabs}), which is sometimes present in other symbiotic systems, but the fit was not improved.
Table \ref{tbl:parameters} summarizes the spectral parameters from the models cited above.

\begin{table}[h!]
\begin{center}
\caption{Best fit spectral parameters from the NuSTAR observation.\label{tbl:parameters}}
\begin{tabular}{lccc}
\tableline\tableline
Parameter &  \multicolumn{2}{c}{\textsc{apec}}   \\
\tableline
$N_H$ (10$^{22}$\,cm$^{-2}$)                     & 19.8$^{+1.8}_{-1.8}$          & 18.3$^{+2.2}_{-2.3}$            \\
$kT$ (keV)                                       & 12.0$^{+0.8}_{-0.7}$          & 9.9$^{+1.1}_{-0.8}$             \\
$Z$ ($Z_{\odot}$)                                & 0.90$^{+0.22}_{-0.19}$        & 0.72$^{+0.17}_{-0.14}$          \\
rel$_{refl}$                                     & ...                           & 1.0$^{+0.8}_{-0.6}$             \\
$\chi^2_{\nu}$/d.o.f.                            & 1.09/182                      & 1.08/181                        \\
F$_{(3-30keV)}^a$  				 & 9.6$\pm$0.8 			 & 9.3$\pm$1.0  	           \\
\hline
 &  \multicolumn{2}{c}{\textsc{mkcflow}} &  \\
\tableline
$N_H$ (10$^{22}$\,cm$^{-2}$)                     & 23.3$^{+2.1}_{-2.3}$          & 22.5$^{+2.4}_{-3.0}$            \\
$kT_{high}$ (keV)                                & 21.1$^{+2.6}_{-1.9}$          & 16.3$^{+4.2}_{-2.7}$            \\
$kT_{low}$ (keV)                                 & 0.0808                        & 0.0808                          \\
$Z$ ($Z_{\odot}$)                                & 0.75$^{+0.20}_{-0.16}$        & 0.59$^{+0.19}_{-0.14}$          \\
rel$_{refl}$                                     & ...                           & 0.9$^{+1.5}_{-0.8}$             \\
$\chi^2_{\nu}$/d.o.f.                            & 1.07/182                      & 1.07/181                        \\
F$_{(3-30keV)}^a$  				 & 10.0$\pm$1.3 		 & 9.9$\pm$1.1                     \\
\hline
\tableline
\end{tabular}
\end{center}
$^a$ Fluxes are unabsorbed and in units of 10$^{-12}$\,erg\,cm$^{-2}$\,s$^{-1}$; \\
The models are M1 [\textsc{tbabs*(apec+gauss)}]
and M2 [\textsc{tbabs*(mkcflow+gauss)}], in the second column, and 
the same but with the thermal component convolved with a \textsc{reflect} component
in the third column.\\
\end{table}

\subsection{Reflection and consequences for the derived plasma properties}
\label{sct:reflect}

Although the inclusion of the \textsc{reflection} component does not improve the $\chi^{2}_{\nu}$ goodness of the fit with respect to ``pure'' models, our spectral fitting indicates that this component could be present in SU\,Lyn. We used the \textsc{reflect} model in \textsc{xspec} in order to check for the presence of a Compton hump that could be due to ``reflection'' of intrinsic X-rays over the white dwarf or nearby cold material. This component was convolved by the thermal components in M1 and M2.
We assumed that the abundance of the \textsc{reflect} component, including the iron abundance, was equal to the abundance of the corresponding thermal component and allowed them to vary while linked during the fit. 

Allowing the reflection scaling factor to vary during the fitting
 resulted in values equal to 0.98$^{+0.84}_{-0.58}$ and 0.87$^{+1.46}_{-0.77}$ from M1 and M2, respectively. 
 The inclination angle $i$ between the normal to the reflector and the line of sight \citep{1995MNRAS.273..837M} was not constrained in the fit, so we set it such that cos($i$) = 0.45.
 With the inclusion of reflection, the best-fit $kT$ value is lower than in the fit without reflection. Although the 1\,$\sigma$ error ranges overlap, the best-fit temperature without reflection (21.1\,keV) is outside the range obtained with reflection (13.6--20.5\,keV), and vice versa (16.3\,keV and 19.2--23.7\,keV; see Table \ref{tbl:parameters}). 
In general, when the statistical quality of the data is higher, the drop in $k$T with the addition of a reflection component is statistically significant \citep[as was the case with RT\,Cru;][]{2018arXiv180102492L}. So we advise caution when the white dwarf mass is derived exclusively using fits without reflection. In this work, we present analyses with and without reflection.

The spectrum of SU\,Lyn is marked by a noticeable excess due to the fluorescent and ionized lines of the Fe\,K complex. We inferred a line intensity of (1.8$\pm$0.3)$\times$$10^{-5}$\,photons\,cm$^{-2}$\,s$^{-1}$ and equivalent width (EW) of 150-380\,eV for the fluorescent Fe line at 6.4\,keV. 
The measured equivalent width does not allow us to distinguish the cases with or without reflection because its lower limit is still consistent with the contribution being only due to the local X-ray absorber \citep[see Figure 7 of][]{1999ApJS..120..277E}, 
while the higher
end of allowed values would appear to require the contribution from both
the absorber and reflection.

\subsection{Photometric and spectral variability}

\subsubsection{X-ray photometric variability}
\label{sect:Xphot}

SU\,Lyn is a variable X-ray/UV source and therefore caution is required when investigating properties from spectra that accumulate information acquired over long time scales -- because the long-term variability is energy dependent. 
However, following the expectation that the plasma temperature -- which strongly depends on the gravitational potential well promoted by the WD -- is not variable, and that the absorption has little effect in hard X-rays, we compare the NuSTAR spectrum with the BAT averaged spectrum constructed by co-adding observations from 2004 December 8 to 2016 January 11, which is the average of ``normal state'' and ``high state'' \citep{2016MNRAS.461L...1M}. Both can be described by the same model being the only noticeable difference associated to the flux that drops during the NuSTAR observation (Section \ref{sect:Xphot}).
The corresponding mean flux from the averaged BAT spectrum was obtained by 
integrating the absorbed APEC model (M1) in a simultaneous fit of the NuSTAR and BAT spectra. 

Although the average BAT spectrum still had a low signal-to-noise, this exercise showed that the NuSTAR flux at the 15-35\,keV energy range in 2016 August (of about 2.5$\times$10$^{-12}$\,erg\,cm$^{-2}$\,s$^{-1}$) was about 47\% of the flux derived from the BAT observations between 2004 and 2016 (of about 5.3$\times$10$^{-12}$\,erg\,cm$^{-2}$\,s$^{-1}$). 
Repeating the exercise with the BAT spectrum corresponding to the time interval associated to the high state of SU\,Lyn described by \citet{2016MNRAS.461L...1M}, we found that the 15-35\,keV flux during the NuSTAR observations in 2016 August corresponds to $\approx$\,14\% of the high state observed with Swift/BAT between 2010 October 14 and 2012 August 1 ($\approx$\,1.8$\times$10$^{-11}$\,erg\,cm$^{-2}$\,s$^{-1}$). In both cases, using the ``averaged'' or ``high state'' BAT spectrum in the simultaneous fit with the NuSTAR spectra, the values obtained for the absorption column, temperature, and abundance are fully consistent with those presented in Table \ref{tbl:parameters}. As presented by \citet{2016MNRAS.461L...1M}, the normal state before and after the high state was about 1/2 of the ``averaged'' flux, hence SU\,Lyn in 2016 August appears to have been at
a similar hard X-ray luminosity level as the normal state.

\begin{figure*}[h!]
\includegraphics[angle=-90,scale=.45]{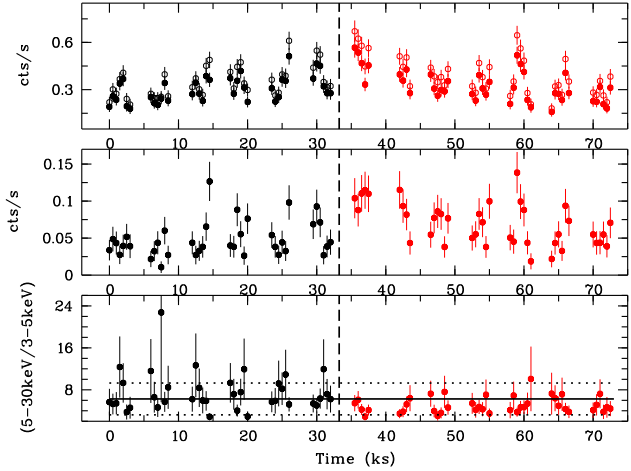}
\caption{NuSTAR X-ray light curves and hardness ratio (FPMA+FPMB). Top: 3--30\,keV (open circles) and 5--30\,keV (filled circles). Middle: 3--5\,keV. Bottom: (5--30\,keV)/(3--5\,keV). The vertical line delimits the first (black) and second (red) halves of the observation explored in Fig. \ref{fig:spct_unf}. The horizontal lines in the bottom panel represent the mean value (solid line) and the corresponding limits at 1$\sigma$. 
The $t_0$ corresponds to 2016 August 12 at 20:25:00.200 UTC, the start time of the NuSTAR exposures. Time bin size is 500\,s.
\label{fig:lcnustar}}
\end{figure*}

Variability on a shorter timescale ($>$\,100\,s) was accessible from the NuSTAR observation. Figure \ref{fig:lcnustar} presents the light curve in the widest possible energy range (3--30\,keV; open circles in the top panel). It also includes two other light curves from the hardest (5--30\,keV; top panel) and softest (3--5\,keV; middle panel) photons and the corresponding hardness ratio (hard/soft; bottom panel). All light curves were constructed by combining the FPMA and FPMB data in time bins of 500\,s. Both soft and hard light curves are marked by variations up to a factor 2.5 on timescales as short as 500\,s. There is marginal evidence from the photometry that SU\,Lyn exhibits spectral evolution with time, especially when comparing the first and second halves of the observation. 
Figure \ref{fig:lcnustar} (bottom) suggests that the hardness ratio is more variable and on average higher during the first half, indicating a higher fraction of hard X-rays or lower fraction of soft X-rays in comparison with the rest of the observation. While the suspicion of fast variability (of about 500\,s) cannot be accessed, with the flux being consistent with the mean value at a 1$\sigma$ level even from the photometry (horizontal lines at the bottom of Fig. \ref{fig:lcnustar}), the variability in 10\,h can be investigated from spectroscopy. We return to latter point in the Section \ref{sct:spctevol}.

\subsubsection{Search for period modulation in X-rays}

\begin{figure*}[h!]
 \centerline{\includegraphics[angle=0,scale=0.52]{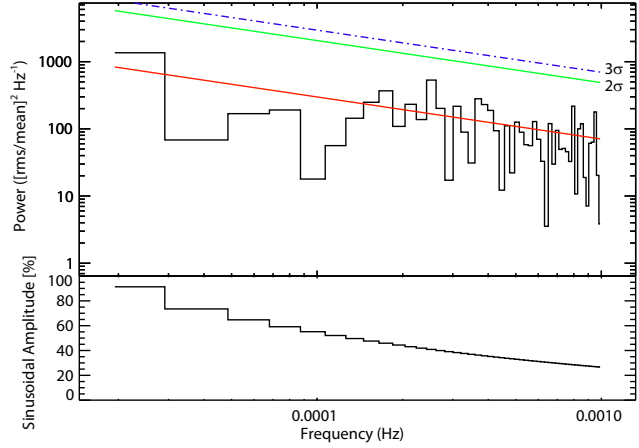}}
 \caption{NuSTAR power spectrum from the 3-30\,keV FPMA+FPMB light curve with bins of 500 s. The power law model for the red
noise is shown in solid (red) line and 2$\sigma$ (green) and 3$\sigma$ (blue) upper limits on the expected power are shown in dotted-dashed curves. Bottom: upper limit (3$\sigma$) on the sinusoidal fractional amplitude as a function of frequency.\label{fig:powerspct}}
 \end{figure*}

We searched for periodic modulation in X-rays by investigating the Fourier power
spectrum of the 3-30\,keV NuSTAR (FPMA+FPMB) light curve binned in 500\,s.
A simple power law model was applied to describe the log-log power spectrum from 10$^{-5}$ to 10$^{-3}$\,Hz and a deviation at a 3$\sigma$ level was obtained by following the method outlined by \citet{2005A&A...431..391V}. Also, the upper limits on the sinusoidal fractional amplitude as a function of frequency were calculated following equation 13 in \citet{1996ApJ...468..369I}.
Since no peak exceeds a 3$\sigma$ detection threshold, there is no evidence for periodic modulation from the NuSTAR data (Fig. \ref{fig:powerspct}).

\subsubsection{Evaluating X-ray spectral evolution}
\label{sct:spctevol}

\begin{figure*}[h!]
\includegraphics[angle=-90,scale=.35]{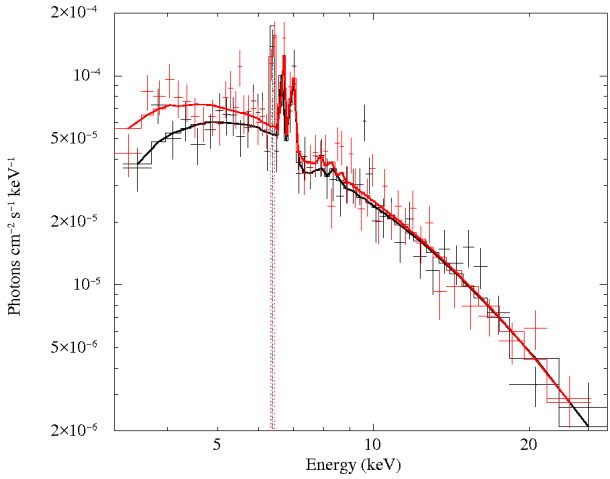}
\caption{NuSTAR unfolded FPMA spectra: black, first half; red, second half (as presented in Fig. \ref{fig:lcnustar} with the same colour code).\label{fig:spct_unf}}
\end{figure*}

Figure \ref{fig:lcnustar} (bottom) suggests that the X-ray hardness ratio of SU\,Lyn was on average higher during the first half of the observation. To evaluate this suspicion, we split the observation into two segments and performed spectral analysis on each segment separately, with a pair of FPMA and FPMB spectra from the first 6 orbits and another pair from the rest (7 orbits). The vertical line in Fig. \ref{fig:lcnustar} shows where we split the data and Fig. \ref{fig:spct_unf} shows the corresponding unfolded spectra (only FPMA, for clarity) -- with both following the same colour code.

We consider two cases using M2 (both without reflection). In the first one, the plasma parameters were frozen to those presented in Table \ref{tbl:parameters} ($k$T\,=\,21.1\,keV and $Z$\,=\,0.75\,$Z_{\odot}$). 
It resulted in 
N$_{H}$\,=\,28.4$^{+2.2}_{-2.1}$\,$\times\,$10$^{22}$\,cm$^{-2}$ for the first half and 
N$_{H}$\,=\,20.6$^{+1.6}_{-1.5}$\,$\times\,$10$^{22}$\,cm$^{-2}$ for the second half of the observation ($\chi^2_{\nu}$/d.o.f\,=\,1.07/89 and $\chi^2_{\nu}$/d.o.f\,=\,1.16/111, respectively). 
Then we considered a second case in which the temperature was free during the fit, resulting in 
N$_{H}$\,=\,28.4$^{+3.0}_{-3.0}$\,$\times$ 10$^{22}$ cm$^{-2}$ and 
$k$T\,=\,21.1$^{+3.0}_{-2.2}$\,keV  with 
$\chi^2_{\nu}$/d.o.f\,=\,1.08/88 for the first half, and 
N$_{H}$\,=\,22.1$^{+2.3}_{-2.2}$\,$\times\,$10$^{22}$\,cm$^{-2}$ and 
$k$T\,= 19.2$^{+2.0}_{-1.8}$ keV with 
$\chi^2_{\nu}$/d.o.f\,=\,1.16/110 for the second one.

This analysis reveals a significant decrease in the local absorption by about 25\% on a timescale of $\sim$\,10 hours, while the properties of the X-ray emitting plasma remained essentially the same, including the unabsorbed flux. Similar results are obtained with M1. 
The impact of the absorption on the continuum can be seen up to 5\,keV (Figure \ref{fig:spct_unf} shows the unfolded spectra for the first case cited above, with parameters frozen).

\subsubsection{UV photometric variability}
\label{sct:uv_variability}

\begin{figure*}[t!]
\centerline{\includegraphics[angle=0,scale=1.]{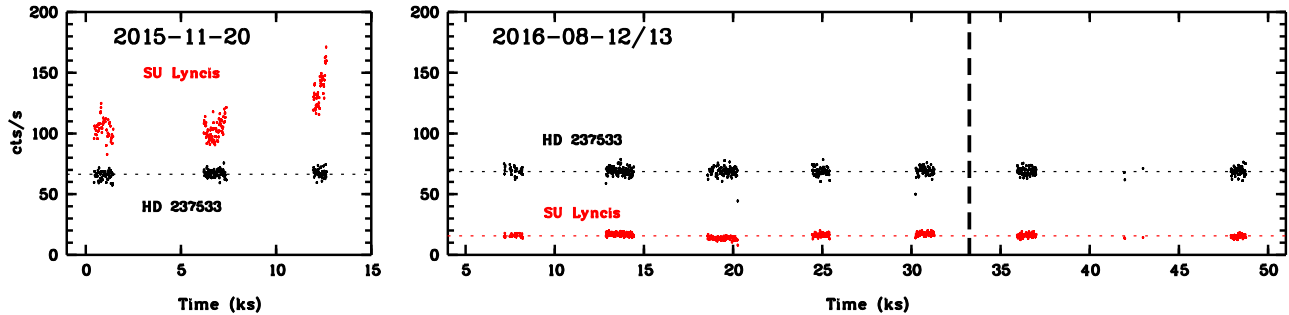}}
\caption{UV light curve (Swift/UVOT; UVM2 filter, centred at 2,246\,\AA). 
The $t_0$ for the first Swift/UVOT exposure (left panel) corresponds to its start time (2015 November 20 at 17:43:48.548 UTC), while the $t_0$ for the second Swift observation (right panel) corresponds to the start time of the NuSTAR exposures (2016 August 12 at 20:25:00.200 UTC) for direct comparison with Fig. \ref{fig:lcnustar}. 
The vertical line in the right panel flags the time used to split the NuSTAR observation into two parts (as in Fig. \ref{fig:lcnustar}). 
Time bin size is 15\,s.
\label{fig:lcuvot}}
\end{figure*}

We compared the recent dataset with that collected five months earlier also with the UVM2 filter, and reported by \citet{2016MNRAS.461L...1M}. Following the procedure adopted by those authors, we use HD\,237533 as a comparison star. Figure \ref{fig:lcuvot} shows the UV light curves of both SU\,Lyn and HD\,237533. While the count rate of the comparison star remained essentially constant during both campaigns (66.40\,$\pm$\,3.40 counts\,s$^{-1}$ and 68.62\,$\pm$\,4.64 counts\,s$^{-1}$ for the whole first and second observations, respectively), the brightness of SU\,Lyn had decreased by a factor of about 8.7 when comparing the mean value registered during the last orbit on 2015 November 20 (137.51\,$\pm$\,13.34 counts\,s$^{-1}$) and that from the orbits on 2016 August 12/13 (15.84\,$\pm$\,1.85 counts\,s$^{-1}$).

Individually, the mean count rates for each orbit of the 2016 observation are consistent at a 1$\sigma$ level with the mean value from the whole light curve. This is also true when dividing the UVOT data into two portions (see the vertical line in Fig. \ref{fig:lcuvot}, which markes the same time as the vertical line in Fig. \ref{fig:lcnustar}) -- with the first portion having been conducted during the first half of the NuSTAR observation and the second portion carried out during the second half of the NuSTAR observation explored in Section \ref{sct:spctevol}. The first portion corresponds to five Swift orbits, with a mean count rate of 15.86$\pm$2.04 counts\,s$^{-1}$, while the second portion was covered by three orbits, from which the mean count rate was 15.58$\pm$1.90 counts\,s$^{-1}$. Thus, contrary to X-rays, the first and second portions of the observation in UV remained essentially the same and were consistent with the mean value for the whole light curve, even with the 25\% drop in the N$_{H}$ affecting the X-rays. However, a difference of about 20\% in count rate can be present from one orbit to another, as seen between the second (17.09$\pm$1.28 counts\,s$^{-1}$) and third (13.49$\pm$1.33 counts\,s$^{-1}$) orbits.

The fractional RMS level of SU Lyn in the Swift/UVOT campaign on 2016 remained essentially constant from one orbit to another, ranging from 7.5 per cent to 9.8 per cent, and about 12 per cent for the whole UV light curve. This is consistent with the values pointed out by \citet{2016MNRAS.461L...1M} for the UVOT campaign carried out in November 2015.

\section{Discussion}

SU Lyn is marked by a hard thermal X-ray emission of moderate luminosity, which is affected by strong and variable local absorption. Both X-ray and UV fluxes vary on short and long timescales. In the following, we discuss the properties of the system as inferred from NuSTAR X-ray spectroscopy and in the light of a dramatic decrease in the L$_{UV}$/L$_{X}$ ratio observed over the course of 9 months.

\subsection{From the X-ray spectroscopy of SU\,Lyn}

 Our analysis showed that both single-temperature (\textsc{apec}) and cooling plasma (\textsc{mkcflow}) models describe well the X-ray spectrum of SU\,Lyn (Table \ref{tbl:parameters}). 
However, a single temperature plasma model is unphysical: thermal emission cools the plasma, for one thing, and a thermal plasma cannot settle onto the white dwarf until it cools to the photospheric temperature, for another. Such a cooling flow is exactly the situation that the \textsc{mkcflow} model represents, with some simplifying assumptions, meaning that this is a more realistic physical model of the X-ray emission from SU\,Lyn \citep[see][]{2017PASP..129f2001M} for a more complete discussion). 
Moreover, direct evidence for a multi-temperature plasma in another symbiotic star with $\delta$-type X-ray emission was provided by the Chandra observation of V658\,Car \citep{2010ApJ...709..816E}.
While a cooling flow can be associated with either an accretion column or a boundary layer in an accretion disk, the lack of periodic X-ray modulations disfavors magnetic accretion and argues in favor of accretion through a boundary layer.

The cooling flow model reveals a maximum temperature $k$T approximately equal to 21\,keV, or 16\,keV if reflection of hard X-rays in cold material is present. 
The data are consistent with either models with or without reflection, as judged by the continuum shape (i.e., whether a Compton reflection hump is present) and by the strength of the fluorescent 6.4 keV line.
With or without reflection, the unabsorbed flux implies a luminosity of 4.9$\times$10$^{32}$ erg\,s$^{-1}$ at the 3--30\,keV energy band, assuming a distance of 640\,pc. Extrapolating the NuSTAR response to 0.1-100\,keV, we estimate the bolometric luminosity to be 7.4$\times$10$^{32}$ erg\,s$^{-1}$ and 9.8$\times$10$^{32}$ erg\,s$^{-1}$ for \textsc{apec} and \textsc{mkcflow} based models, respectively.

The maximum temperature in the boundary layer suggests that the mass of the white dwarf is at least about 0.8\, M$_{\odot}$.
In non-magnetic accreting systems in which an accretion disk is formed, approximately half the potential energy of the falling matter is radiated away in the disk where the Keplerian velocity corresponds to (1/2)$^2$ times free-fall velocity. Assuming that this is the modus operandi for SU\,Lyn, the maximum shock temperature (T$_{s,max}$) comes from the X-ray spectral fits using the \textsc{mkcflow} model (M2 in Section \ref{sct:spct}) and allows us to estimate the white dwarf mass (M$_{WD}$). 
Doing this, we calculate half of the total energy from free-fall from infinity following \citet{2002apa..book.....F}, assuming the mass-radius relation for white dwarfs suggested by \citet{1975MNRAS.172..493P}. 
Factors such as rapid rotation, high core temperature, and high envelope temperature can all modify the mass-radius relationship in principle. However, they all go in the direction of larger radii, and hence the lower limits we derive below on the white dwarf mass
and the mass accretion rate are secure.
This theoretical approach was successfully applied by \citet{2010MNRAS.408.2298B} to explain the locus occupied by a sample of dwarf novae in the M$_{WD}$\,versus\,T$_{s,max}$ diagram. Our results indicate that the mass of the white dwarf in SU\,Lyn is consistent with 
0.87$^{+0.05}_{-0.04}$\,M$_{\odot}$ for T$_{s,max}$\,=\,21.1$^{+2.6}_{-1.9}$\,keV (from the \textsc{mkcflow} model), 
and can be 14\% less, 
0.76$^{+0.10}_{-0.07}$\,M$_{\odot}$ 
if reflection is present (Section \ref{sct:reflect}).
From this model, the mass accretion through the optically thin portion of the boundary layer is approximately equal to (1.8$\pm$0.3)$\times$10$^{-10}$ M$_{\odot}$\,yr$^{-1}$, or (2.0$\pm$0.2)$\times$10$^{-10}$ M$_{\odot}$\,yr$^{-1}$ if reflection is present. When the UV luminosity exceeds the X-ray luminosity, part of the boundary layer is likely optically thick, and therefore the values above correspond to lower limits for the mass accretion rate.

\subsection{From the spectral and photometric variability}

Our observations show that in 2016 the UV emission from SU Lyn had dropped after the high state in 2015 reported by \citet{2016MNRAS.461L...1M}. These authors reported from Swift/UVOT and GALEX data that SU Lyn is variable in UV on timescales as short as tens of seconds with a fractional variability of 7--10 per cent. 
They reported from GALEX data that SU\,Lyn was much fainter in UV on 2006 December 21 and 2007 January 27 than Swift/UVOT showed it to be on 2015 November 20.
In the Swift/UVOT data, \citet{2016MNRAS.461L...1M} also observed that the UVM2 count rate increased by about 30 per cent within 95 minutes from the second to the third and last orbit of the campaign on 2015 November 20 (Fig. \ref{fig:lcuvot}, left). 

We show here that SU\,Lyn underwent a dramatic decrease by a factor 8.7 in the UV flux on an unknown timescale but which is constrained to be less than 9 months, the elapsed time between the two recent Swift/UVOT observations (2016-08-12 and 2015-11-20; Fig. \ref{fig:lcuvot}). The low state lasted for at least 11.6 hours. We refer to this behavior as ``long-term variability'', if not for the timescale on which it occurred, then for the duration of the states. 
It corresponds to a change from 
(1.16$\pm$0.11)$\times$10$^{-13}$\,erg\,s$^{-1}$\,cm$^{-2}$\,\AA$^{-1}$, to (1.34$\pm$0.16)$\times$10$^{-14}$\,erg\,s$^{-1}$\,cm$^{-2}$\,\AA$^{-1}$, in the spectral range covered by the UVM2 filter, and therefore to a drop in the total UV at 2000--4000\AA\, from 
2.3$\times$10$^{-10}$\,erg\,s$^{-1}$\,cm$^{-2}$
to 
2.7$\times$10$^{-11}$\,erg\,s$^{-1}$\,cm$^{-2}$. 
In terms of luminosity, these values correspond to  
1.1$\times$10$^{34}$($d$/640 pc)$^2$\,erg\,s$^{-1}$
and
1.3$\times$10$^{33}$($d$/640 pc)$^2$\,erg\,s$^{-1}$,
respectively.
In contrast, the hard X-ray flux observed with NuSTAR in 2016 in
the 15-35 keV energy band is similar to that seen with Swift/BAT
during the normal state, as defined by \citet{2016MNRAS.461L...1M}.

The intervening column affecting X-rays changed significantly from 28$\times$10$^{22}$\,cm$^{-2}$ to 21$\times$10$^{22}$\,cm$^{-2}$ in the course of the NuSTAR observation (Section \ref{sct:spctevol}), proving that changes may happen on timescales as short as 10\,ks. 
It suggests that the spikes in the hardness ratio (Fig. \ref{fig:lcnustar}; bottom) are associated with a decrease of softest X-rays (3--5\,keV) caused by an increase of the photoelectric absorption that, on average, results in a higher column for the first half than that for the second half of the observation. In this case, the ``instantaneous'' absorption in the short period of time associated with the ``hardness spikes'' may be even higher than the inferred value. What could be causing this variation? Despite the complexity in the absorption with time, the X-ray spectrum (3--30\,keV) is well described for a simple absorption component. Although this can only be confirmed by sensitive observations including softest X-rays, the results suggest that the intervening material may be inhomogeneous but relatively well distributed over the X-ray emitting sites.

As for SU\,Lyn, there is also evidence that X-rays of $\delta$-type symbiotics as a whole suffer the effect of local absorbers that are variable on a day to day time scales. 
Regardless of whether this is due to spatial inhomogeneities or due to temporal changes, the timescale of
the N$_{H}$ variability can be used to constrain the origin of the absorber. In particular, it is unlikely to be the wind of the mass donor: the binary likely has a scale of $\sim$\,AU, and the wind of a red giant has a characteristic velocity of order 10 km\,s$^{-1}$, so it would be difficult for this to lead to variable N$_{H}$ on timescales much shorter than 1 AU / 10 km\,s$^{-1}$ $\approx$\,6 months. An origin much closer to the white dwarf is indicated.

The local X-ray absorber does not appear to absorb the UV emission from SU\,Lyn,
as evidenced by the lack of detectable changes in UV count rates between the first
and second halves of the NuSTAR observation, when the local X-ray absorber varied
significantly.
We can understand this in the context of localized absorbers in two possible ways. If the absorber is extremely localized, right next to the X-ray emission region (presumably the boundary layer
between the disk and the white dwarf), then it might not obstruct our view of the UV emission region (parts of the Keplerian accretion disk proper). Alternatively, the X-ray absorber might cover both the X-ray and UV emitting regions but may be transparent to UV. This is possible because UV absorption in the ISM
is due to dust and molecules, while that in interacting binaries may be due to
~10,000K material \citep[``Fe II Curtain'';][]{1994ApJ...426..294H} and the localized absorber
in SU Lyn may well be too hot for either.

\subsection{The scenario}

Whereas the Keplerian part of the accretion disk accounts for the UV radiation, the boundary layer accounts for the X-rays. \citet{2016MNRAS.461L...1M} suspected that the boundary layer during the Swift observation on 2015 November 20 was at least partially optically thick to X-rays -- implying that the WD mass estimate they obtained from X-rays, 1\,M$_{\odot}$, is a lower limit. 
Changes in the ratio of L$_{UV}$ to L$_{X}$ give us clues about the physical conditions in the boundary layer. Assuming that the observed UV is not subject to strong intrinsic absorption, as is the case for the X-ray photons in the 15-35\,keV energy regime, we use the L$_{UV}$/L$_{X;15-35\,keV}$ as a proxy of the boundary layer conditions. 
The NuSTAR flux was similar to that during the normal state as seen by BAT \citep[2004 December 8
through 2010 October 13, and 2012 August 2 to 2016 January 11; see Figure 1 of][]{2016MNRAS.461L...1M}.
Since the 2015 November Swift observations took place during the normal state, the instantaneous
X-ray flux at that time was probably similar to the NuSTAR measurement; however, we cannot be
certain due to the optical loading issue affecting the XRT data.

Therefore L$_{X;15-35\,keV}$ was formally constant ~(about 1.2$\times$10$^{32}$($d$/640\,pc)$^{2}$\,erg\,s$^{-1}$) while L$_{UV}$ dropped from \,1.1$\times$ 10$^{34}$($d$/640\,pc)$^{2}$\,erg\,s$^{-1}$ to 1.3$\times$10$^{33}$($d$/640\,pc)$^{2}$\,erg\,s$^{-1}$, with L$_{UV}$/L$_{X;15-35\,keV}$ changing from 84 to 11. 
Even in a narrow band  (2000--4000\,\AA) the UV luminosity exceeds the estimated bolometric X-ray luminosity (estimated to be 9.8$\times$10$^{32}$($d$/640\,pc)$^{2}$\,erg\,s$^{-1}$), whereas in the case of an optically thin boundary layer, it is expected that X-ray emission roughly equals the bolometric disk luminosity. These UV and X-ray features suggest: 
(i) a substantial decrease in the total mass accretion rate and 
(ii) that more of the
  boundary layer became optically thin to X-ray photons, so that any
  decrease in X-ray luminosity that we might have expected from the
  drop in accretion rate was compensated for by the increased fraction
  of the boundary layer emitting in the X-ray regime.

\section{Summary}

The main findings of this paper came from the first reliable X-ray spectroscopy of SU Lyn and complementary UV photometry. They are:
\begin{enumerate}
     \item The hard X-ray spectrum is consistent with the presence of reflection of hard X-rays from cold material, with reflection amplitude (R) equal to 1.
     \item We revised the WD mass estimate from \citet{2016MNRAS.461L...1M} taking into account the effects of the reflection component, with R\,=\,1 and R\,=\,0 fits resulting in a minimum mass between about 0.7 and 0.8\,M$_{\odot}$, respectively.
     \item We identified strong and variable intrinsic X-ray absorption, with rapid variability suggesting that the absorber is near the accreting white dwarf.
     \item The X-ray absorber appears not to absorb UV. This implies that the absorber is extremely localized, predominantly in the line of sight of the primary X-ray emitter, and/or significantly ionized, and therefore without molecules and dust that could affect the UV photons.
     \item Between 2015 November and 2016 August, the L$_{UV}$/L$_{X}$ ratio dropped dramatically,
      supporting a decrease in accretion rate while the boundary layer became more optically thin.
 \end{enumerate}

We encourage further observations of SU Lyn to refine our findings and to take further
advantage of its long-term variability to study the response of accretion disk and the boundary layer
to changes in accretion rate.

\acknowledgments
Acknowledgments.

R.L.O. was partially supported by the Brazilian agency CNPq (PDE 200289/2017-9, Universal Grants 459553/2014-3, PQ 302037/2015-2).
J. L. S. acknowledges support from NASA grants NNX15AF19G and NNX17AC45G.
G.J.M.L. is a member of the CIC-CONICET (Argentina) and acknowledges support from grant ANPCYT-PICT 0478/14 and PIP-CONICET/2011 \#D4598.
The authors acknowledge Katja Pottschmidt and Brian Grefenstette for their guidance on reduction of NuSTAR data.
This research has made use of the NuSTAR Data Analysis Software (NuSTARDAS) jointly developed by the ASI Science Data Center (ASDC, Italy) and the California Institute of Technology (Caltech, USA). This research has made use of the XRT Data Analysis Software (XRTDAS) developed under the responsibility of the ASI Science Data Center (ASDC), Italy. This work made use of data supplied by the UK Swift Science Data Centre at the University of Leicester. 
This research has made use of data, software and web tools obtained from the High Energy Astrophysics Science Archive Research Center (HEASARC), a service of the Astrophysics Science Division at NASA/GSFC and of the Smithsonian Astrophysical Observatory's High Energy Astrophysics Division.
This work has made use of data from the European Space Agency (ESA) mission
{\it Gaia} (\url{https://www.cosmos.esa.int/gaia}), processed by the {\it Gaia}
Data Processing and Analysis Consortium (DPAC,
\url{https://www.cosmos.esa.int/web/gaia/dpac/consortium}). Funding for the DPAC
has been provided by national institutions, in particular the institutions
participating in the {\it Gaia} Multilateral Agreement.

\facilities{Swift (XRT, BAT, and UVOT), NuSTAR (FPMA and FPMB)}

\end{document}